\documentclass{article}

\pdfoutput=1 
\usepackage{booktabs}
\usepackage{multirow}
\usepackage{graphicx}
\usepackage[table,xcdraw]{xcolor}
\usepackage{amsmath}
\usepackage{amssymb}
\usepackage{float}
\usepackage{todonotes}

\usepackage{mdframed}
\usepackage{rotating}
\usepackage{enumitem}
\usepackage{microtype}
\usepackage{lipsum}
\usepackage[american]{babel}
\usepackage[affil-it]{authblk}
\usepackage{hyperref}
\usepackage[numbers]{natbib}
\usepackage[T1]{fontenc}
\usepackage{textcomp}
\usepackage{lmodern}
\usepackage{afterpage}

\newcommand{\fig}[1]{Fig.~\ref{fig:#1}}
\newcommand{\tion}[1]{\S\ref{sec:#1}}

\newcommand{\tab}[1]{Table~\ref{table:#1}}

\newcommand{\bi}{\begin{itemize}}
\newcommand{\ei}{\end{itemize}}
\newcommand{\be}{\begin{enumerate}}
\newcommand{\ee}{\end{enumerate}}
\newcolumntype{L}[1]{>{\raggedright\let\newline\\\arraybackslash\hspace{0pt}}m{#1}}
\newcolumntype{C}[1]{>{\centering\let\newline\\\arraybackslash\hspace{0pt}}m{#1}}
\newcolumntype{R}[1]{>{\raggedleft\let\newline\\\arraybackslash\hspace{0pt}}m{#1}}

\usepackage [autostyle, english = american]{csquotes}
\MakeOuterQuote{"}

\usepackage{tablefootnote}
\usepackage{caption} 
\usepackage{nowidow}

\begin{document}
	
	\title{System-of-Systems Viewpoint for System Architecture Documentation}
	\date{Revised 11 Nov 2017}
	
	\author{John Klein\thanks{john.klein@computer.org, Corresponding author} 	}
	\author{Hans van Vliet\thanks{hans@cs.vu.nl}}
	\affil{VU University, Amsterdam, Netherlands}
	\maketitle
	
\begin{abstract}
\noindent\textbf{Context}: The systems comprising a system of systems (SoS) are independently acquired, operated, and managed. Frequently, the architecture documentation of these existing systems addresses only a stand-alone perspective, and must be augmented to address concerns that arise in the integrated SoS.\\
\textbf{Objective}: We evaluated an architecture documentation viewpoint to address the concerns of a SoS architect about a constituent system, to support SoS design and analysis involving that constituent system.\\
\textbf{Method}: We performed an expert review of documentation produced by applying the viewpoint to a system, using the active review method.\\
\textbf{Results}: The expert panel was able to used a view constructed using the baseline version of the viewpoint to answer questions related to all SoS architect concerns about a constituent system, except for questions concerning the interaction of the constituent system with the platform and network infrastructure.\\
\textbf{Conclusions}: We found that the expert panel was unable to answer certain questions because the baseline version of the viewpoint had a gap in coverage related to relationship of software units of execution (e.g., processes or services) to computers and networks. The viewpoint was revised to add a Deployment Model to address these concerns, and is included in an appendix.
\end{abstract}


\section{Introduction} \label{sec:vp-intro}
A system of systems (SoS) is created by composing constituent systems. Each constituent system retains operational independence (it operates to achieve a useful purpose independent of its participation in the SoS) and managerial independence (it is managed and evol\-ved, at least in part, to achieve its own goals rather than the SoS goals) \cite{Maier-Principles}. In order to assess suitability of the system for use in the SoS and to reason about SoS functionality and quality attributes, the architect of a SoS relies on documentation about the constituent system. In an ideal world, constituent system documentation would be available and address all SoS concerns. Our previous research (discussed in Related Work below) reports that this is not usually the case \cite{Klein-Survey}. The challenge of documenting architectures whose parts are designed by separate organizations is a fundamental challenge of SoS and ultra-large scale systems \cite{Northrop:2006}.

Pragmatically, the SoS architect seeking information about a constituent system has limited options. If there is documentation and/or source code available, the SoS architect can attempt to learn enough about the constituent system design to address concerns about how that system will operate in the SoS. However, constituent systems are developed independently, and often there is limited or no access to documentation or code.  The architect of the SoS could seek to collaborate with the architect of each constituent system to augment the constituent system architecture documentation with the information needed to address the SoS concerns. However, the managerial independence of the development and evolution of constituent systems within a SoS \cite{Maier-Principles} often creates barriers to collaboration. Consider three examples of these barriers:
\be
\item Each constituent system owner retains independent management of funding and objectives, and the constituent system architect's responsibilities for delivering system-oriented capabilities may not provide slack time to allow collaboration with the SoS architect. 
\item There is no ongoing development on a particular constituent system, and so there is no architect assigned who could collaborate with the SoS architect. 
\item Firms are integrating IT systems after a merger or acquisition, and the architects of particular acquired systems have been reassigned or dismissed, and so are not available to collaborate.
\ee

In each of these scenarios, collaboration between the SoS architect and the architects of each constituent system may become a tightly planned and managed, high ceremony event. The SoS architect must articulate a precise request for information, for which the constituent system architect estimates the cost to respond. The SoS owner and the constituent system owner negotiate to fund the constituent system architect's work to respond, and eventually the constituent system architect is directed to supply the requested information to the SoS architect. There is often little or no ability to iterate the information requests or seek elaboration of the responses, and given these high stakes, the architect of the SoS needs a pedigreed basis for a request for information.

The contribution of this paper is an architecture documentation viewpoint to assist SoS architects in collecting or creating sufficient documentation about constituent systems in a SoS. The viewpoint addresses stakeholder concerns about SoS design and analysis. This reusable "library viewpoint" conforms to the ISO/IEC 42010 standard for architecture description \cite{ISO42010}, and provides guidance for SoS architects to request sufficient information about constituent system architectures to satisfy SoS-level concerns about each constituent system operating in the SoS context. The viewpoint was evaluated by an expert panel in a single case mechanism experiment using the active design review method \cite{Parnas:1985}. We found that the baseline version of the viewpoint covered most SoS stakeholder concerns; however, the experiment uncovered a gap  in the area of deployment of software units to computers and networks. We describe how the viewpoint was reworked by adding a new model kind to address this gap. 

This paper is organized as follows: \tion{vp-related-work} discusses related work in the areas of SoS and architecture documentation. \tion{vp-approach} describes our approach to developing and evaluating the viewpoint, which was based on Wieringa's design cycle \cite{Wieringa:2014}. \tion{vp-results} presents the results of our evaluation experiment, and our analysis and interpretation, including how the baseline version of the viewpoint was reworked based on the results of the experiment.  \tion{vp-conclusions} summarizes our conclusions, and the reworked viewpoint is included as an appendix.

\section{Related Work} \label{sos-viewpoint-related-work} \label{sec:vp-related-work}
Generally, concern-driven architecture documentation approaches organize architecture documentation into \emph{views} to address stakeholder concerns \cite{ISO42010,Clements:2011,RozanskiWoods:2005}. These approaches are widely used for software system architecture documentation, for example in the Rational Unified Process (RUP) 4+1 Views \cite{Kruchten:2003}.

At the SoS level, view-based frameworks such as DoDAF \cite{DOD:2010} and MODAF \cite{MOD:2012} have emerged to document SoS architecture. The EU COMPASS Project \cite{COMPASS:2014}, which ended in 2014, addressed SoS modeling, and SoS architecture documentation continues to be an area of active research \cite{Guessi:2015}, producing new documentation approaches such as S3 \cite{Brondum:2010} and SySML-based approaches from the EU AMADEOS project \cite{Mori:2017}.

The architect of a SoS must depend on the documentation of constituent systems. Our earlier research reported that one challenge to designing a SoS architecture is gaps in the architecture documentation of the constituent systems \cite{Klein-Survey}. The architecture documentation of each constituent system usually focuses on the stand-alone operation of that system, and on the stand-alone development and evolution of that system. The architecture documentation for constituent systems is created during engineering development of the constituent system, for different purposes than SoS, notably constituent system bounds, constituent system goals, different modeling goals, and different characteristics of interest \cite{Honour:2013}. While functional interaction with external systems in support of the system's standalone operation may be addressed by the architecture documentation, the quality attribute aspects of those interactions are typically not well covered. 

Participation of a constituent system in a SoS introduces new concerns about that system; however, a survey by Bianchi and colleagues found that there are no applicable quality attribute frameworks for these concerns \cite{Bianchi:2015}. Our earlier research found interoperability to be a primary concern of SoS designers \cite{Klein-SLR}, and more recent work by Batista \cite{Batista:2013} and by Guessi and colleagues \cite{Guessi:2015} confirmed that interoperability is a primary focus of SoS architecture documentation. The system mission characterization of Silva and colleagues provides insight into functional interoperation concerns \cite{Silva:2014}. Architecture documentation for constituent systems that addresses standalone operation may not address SoS interoperability concerns, which go beyond interface syntax. As the context for interface semantics is expanded to the SoS, behavior that might have been considered private to the system becomes externally visible. For example, design decisions such as whether to retry a failed request to an external system may not be architectural in the context of standalone operation, but become externally visible and architectural in the context of SoS operation.

The viewpoint that we developed could be considered an extension of the System Context Viewpoint defined by Woods and Rozanski \cite{Woods:2009}, or of the system context diagram in the \emph{Beyond Views} section of a Views and Beyond architecture document \cite{Clements:2011}. However, each of these focuses on how external interfaces and interactions support the independent operation of the system, and not on how the system interoperates with other systems to achieve an SoS capability. 

\section{Approach} \label{sec:vp-approach}
Our objective is to design an artifact that contributes to the achievement of some goal. Here, the artifact is an architecture viewpoint, and the goal is to allow SoS architects to reason about a constituent system to design a SoS. Wieringa labels this a \emph{design problem} and we used the \emph{Design Cycle} approach \cite{Wieringa:2014} as follows:
\be
\item Problem Investigation---We built on the related work discussed above to identify stakeholders in the SoS design process and their concerns related to constituent systems operating in the SoS context. 
\item Treatment Design---We defined an architecture viewpoint to address those stakeholder concerns. 
\item Treatment Evaluation---We evaluated the treatment by a single case mechanism experiment \cite{Wieringa:2014}, using an expert panel to conduct an active design review \cite{Parnas:1985}. 
\ee

\subsection[Problem Investigation---Identify Stakeholders and Concerns]{Problem Investigation---Identify \newline Stakeholders and Concerns}
There are many stakeholders in a SoS and in its architecture \cite{Bergey:2009}. Our focus is on the architecture design and analysis task, and specifically, reasoning about a constituent system in the context of the SoS, which narrowed the scope to the stakeholder roles listed in \tab{vp-table1}.

\begin{table}[ht!]
	\caption{Selected SoS Architecture Stakeholders}
	\label{table:vp-table1}
	\centering 
	\small
	\begin{tabular}{p{4cm} p{6cm}}
		\toprule
		\textbf{Stakeholder Name}   & \textbf{Stakeholder Role} \\ \midrule
		SoS Architect               & Creates architecture designs to allow constituent systems to interoperate to achieve SoS goals. Proposes or defines necessary or desirable changes to constituent systems.  \\ \midrule
		SoS Program Manager         & Has ultimate responsibility for achieving SoS goals. Negotiates with program managers of constituent systems to make necessary or desirable changes to constituent systems. \\ \midrule
		Developer                   & Makes necessary or desirable changes to the software of the constituent systems.\\ \midrule
		SoS Testers and Integrators & Installs, configures, and tests the constituent systems interoperating as a SoS.\\ \bottomrule
	\end{tabular}
\end{table}

These stakeholders were selected because they are directly involved in understanding the constituent system architectures, proposing or defining changes to those architectures for use in the SoS, and then constructing, testing, and integrating the constituent systems in the SoS.

Our earlier systematic review found that SoS research has heavily focused on interoperability concerns \cite{Klein-SLR}, however, our state of the practice survey indicated that practitioners designing and analyzing SoS architectures have broader technical and non-technical concerns \cite{Klein-Survey}. Since this treatment will be employed by practitioners, we decided to augment the researcher-oriented findings with a survey of practitioner-focused literature to identify additional concerns about constituent systems when designing and analyzing SoS architectures. 

The survey focused on an annual practitioner conference organized by the Systems Engineering Division of the National Defense Industry Association (NDIA)\footnote{See \url{http://www.ndia.org/divisions/systems-engineering}}. We reviewed all papers in the SoS Track and Architecture Track for the conferences from 2009 through 2016, and identified 14 papers that discussed SoS architecture concerns. We also reviewed the United States Department of Defense Systems Engineering Handbook for Systems of Systems \cite{DOD-SEforSoS}, which provides guidance to a broad community of practice. From these sources, we identified a set of concerns that are shown in \tab{vp-table2}. As described below, these concerns were used to define the architecture viewpoint artifact.

\begin{table}[ht!]
	\caption{SoS Stakeholder Concerns from Practitioner-oriented Literature}
	\label{table:vp-table2}
	\centering 
	\small
	\begin{tabular}{p{3cm} p{8cm}}
		\toprule
		\textbf{Publication} & \textbf{Concerns about constituent systems in an SoS}\\ \midrule
		Benipayo 2016 \cite{Benipayo:2016} & Dependencies on other constituent systems \\ \midrule
		Sitterle 2016 \cite{Sitterle:2016} & Interface adaptability (Interoperability), Recovery\\ \midrule
		Gump 2014 \cite{Gump:2014} & Interoperability, Dependencies on other constituent systems\\ \midrule
		Manas 2014 \cite{Manas:2014} & Dependencies on other constituent systems, Shared Resources\\ \midrule
		Carson 2014 \cite{Carson:2014} & Dependencies on other constituent systems\\ \midrule
		Guertin 2014 \cite{Guertin:2014} & Portability, Scalability, Dependencies on other constituent systems, Security\\ \midrule
		Baldwin 2014 \cite{Baldwin:2014} & Stakeholders, Dependencies on other constituent systems\\ \midrule
		Gagliardi 2013 \cite{Gagliardi:2013} & Shared resources\\ \midrule
		Pritchett 2013 \cite{Pritchett:2013} & Dependencies on other constituent systems\\ \midrule
		Dahmann 2012 \cite{Dahmann:2012a} & Interoperability\newline 
		Perceived needs of constituent systems\newline Processes, cultures, working practices between different participating organizations\newline Dependencies at development time and run time\\ \midrule
		Dahmann 2012 \cite{Dahmann:2012}  & Dependencies on other constituent systems\\ \midrule
		Smith 2011 \cite{Smith:2011}      & Interoperability context: assumptions, constraints, drivers\\ \midrule
		Lane 2010 \cite{Lane:2010}        & Monitoring and measurement \\ \midrule
		DoD 2008 \cite{DOD-SEforSoS}      & Technical and organizational dependencies\newline Interoperability\newline
		Synchronization of delivery of features across constituent systems (dependencies)\newline Constituent system stakeholders\newline Constituent system needs and constraints\newline Constituent system evolution strategy and built-in variabilities \\ \bottomrule
	\end{tabular}
\end{table}

\subsection{Treatment Design---Define the Architecture Viewpoint}
Wieringa defines a \emph{treatment} as "the interaction between the artifact and the problem context" \cite[\S3.1.1]{Wieringa:2014}. We will define an artifact---an architecture viewpoint---that will be applied by an SoS architect to create an architecture view of a constituent system that provides the information needed to reason about that constituent system when it is operating in the context of the SoS. In this section we focus on the design of the architecture viewpoint, however, our evaluation will consider the entire treatment.

The viewpoint definition conforms to the ISO/IEC/IEEE 42010 standard \cite{ISO42010}, using Annex B of that standard as the template for the viewpoint specification. This approach was selected because of its status as a global standard and because it is compatible with producing documentation using other approaches such as Views and Beyond (see, for example, Appendix E in \cite{Clements:2011}).

\begin{figure}[tb]
	\centering 
	\small
	\includegraphics[width=.9\textwidth]{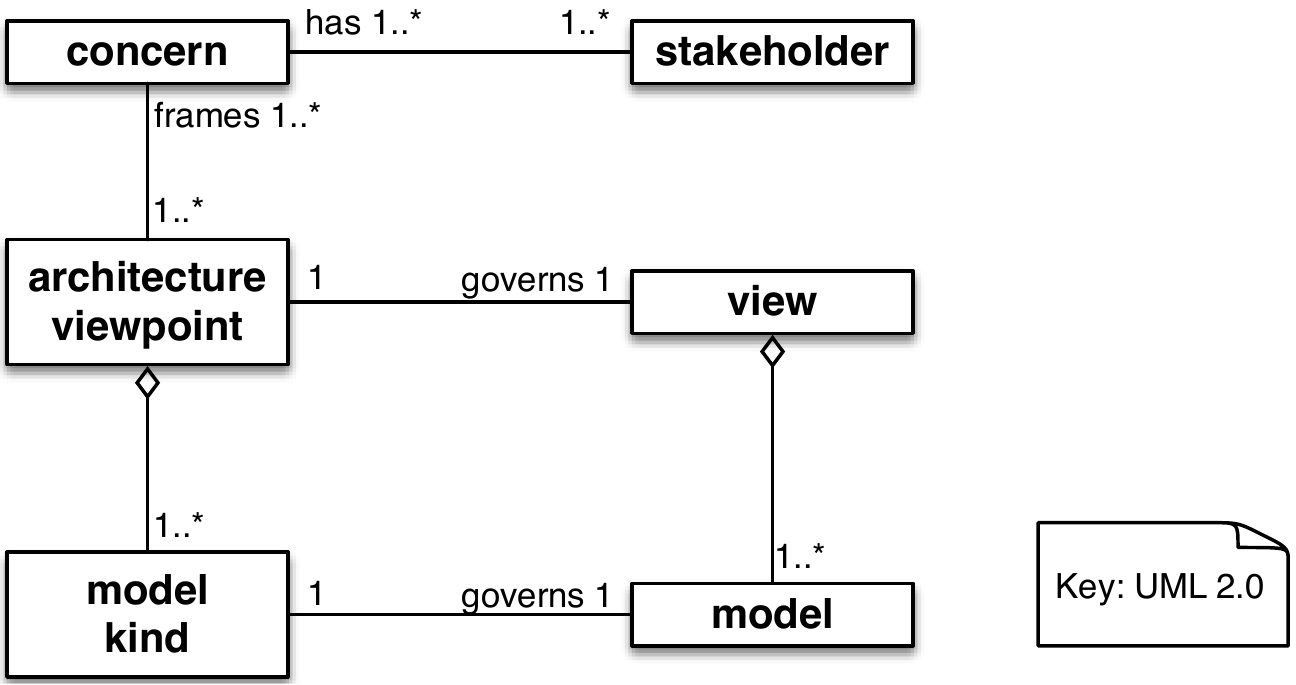}
	\caption{Extract from ISO 42010 Conceptual Model (Adapted from \cite{ISO42010})}
	\label{fig:vp-fig1}
\end{figure}

The subset of the ISO 42010 conceptual model related to viewpoint definition is reproduced in \fig{vp-fig1}. 

As shown in \fig{vp-fig1}, the viewpoint definition begins by identifying stakeholders and concerns. The concerns identified in \tab{vp-table2} are somewhat general. In order to define an architecture viewpoint to address the concerns, we refined these by mapping them to the set of quality attributes that Bass and colleagues \cite{Bass-SAiP} defined and found to be relevant to all software systems, namely performance, availability, security, testability, modifiability, and usability. In \tab{vp-table3}, we consider each of these quality attributes (along with a category for concerns about the system context that are shared by many stakeholders) as concerns at the SoS level, and then trace down to information needed at the constituent system level in order to address the SoS concern. This tracing was performed by considering the tactics \cite{Bass-SAiP} that might be applied to achieved the quality. Tactics that could be applied in the SoS context became concerns about constituent systems in \tab{vp-table3}. Bass and colleagues also discuss which stakeholders are typically concerned with each quality attribute, and we include this information in \tab{vp-table3}. Note that the SoS architect is concerned with all qualities.
\noclub[3]
\begin{table}[ht!]
	\caption{Tracing SoS Concerns into Constituent Systems}
	\label{table:vp-table3}
	\centering 
	\footnotesize
	\begin{tabular}{p{1.75cm} p{6.5cm} p{2.25cm} } \toprule
		\textbf{SoS\newline Concern} & \textbf{Constituent System Concern} & \textbf{Stakeholder} \\ \midrule 
		Performance & 
		Shared resources: what is shared, how is use shared, behavior when insufficient resource is available (run time dependencies, monitoring and measurement, interoperability context) & 
		SoS Architect\newline Program Manager\\ \midrule 
		Security & 
		Authentication: identity validation repository (interoperability)\newline 
		Authorization: remote access to system and resources (interoperability)\newline
		Encryption: algorithms and key management (interoperability) & 
		SoS Architect\newline Tester/Integrator\newline Program Manager \\ \midrule 
		Testability & 
		Execution time dependencies: startup sequencing (run time dependencies)\newline Fault detection and logging: internal (monitoring and measurement) & 
		SoS Architect\newline Tester/Integrator \\ \midrule 
		Modifiability & 
		Build time dependencies: COTS, FOSS, development environment, process/culture/working practices\newline Run time dependencies on other constituent systems\newline Variabilities: Affecting interfaces, decision model (dependencies) (evolution and built-in variabilities) & 
		SoS Architect\newline Developer\\ \midrule 
		Availability & 
		Fault detection and logging: interfaces (monitoring and measurement)\newline Fault recovery (interoperability context) &
		SoS Architect\newline Tester/Integrator\newline Program Manager \\ \midrule 
		Usability (for SoS operators) & 
		Configuration dependencies among constituent systems (development time and run time dependencies) & 
		SoS Architect\newline \\ \midrule 
		Context & 
		Perceived needs and constraints of constituent systems\newline 
		Processes, cultures, working practices between different participating organizations\newline
		Constituent system stakeholders & 
		All\\ \midrule 
	\end{tabular}
\end{table}

\fig{vp-fig1} shows that the viewpoint comprises one or more model kinds. The model kinds were developed iteratively, using the following approach:
\be
\item Identify the type of elements and relations needed to address each concern.
\item Group concerns that had the same types of elements and relations. 
\item Define a model kind for each group of concerns.
\ee

Following this approach, we developed five model kinds, each addressing particular concerns from  \tab{vp-table3}, and collectively addressing all concerns. The model kinds are listed below. Each model kind includes a brief discussion of the model elements and relationships, and the complete definition of the model kinds is provided in the appendix:
\bi
\item Constituent System Stakeholders/Concerns--provides architecture context for the SoS architect and Program Manager by providing insight into the perceived need of the constituent system, and identifies stakeholders who may be impacted by the constituent system's operation within the SoS. The model elements are constituent system stakeholders and stakeholder concerns, with the relation \emph{stakeholder has a concern}.
\item Constituent System Execution Time Context---addresses concerns related to dependencies at execution time, shared resources, and to a lesser extent, provides overall context. The model elements are the constituent system and external software that the system interacts with during execution. The relations are execution time interactions, e.g., sends/receives message, call/return, read/write data, etc.
\item Constituent System Code Context---addresses concerns related to implementation dependencies. The model elements are the constituent system software, and external modules (e.g., libraries, development tools, packages, etc.). The relation is \emph{uses}.
\item Constituent System Interface Information Model---addresses concerns related to semantic interoperation of data elements. The elements are information elements of interest to the SoS (e.g., a SoS that deals with geo-location might use concepts like position, elevation, and direction), and information elements in the constituent system software. Relations are \emph{logical associations} (1-to-1, 1-to-N, N-to-M), \emph{specialization/generalization}, and \emph{aggregation}.
\item Shared Resource Model---addresses concerns about runtime resource sharing. Elements are components in the SoS representing resources used by the constituent system and by other systems, including processor compute cycles, memory, disk space, network bandwidth, files, databases, virtual infrastructure, or physical resources such as a display, antenna, or radio frequency. Relations are \emph{acquires/releases} and \emph{consumes}.
\ei

\tab{vp-table4} shows the mapping from the concerns listed in \tab{vp-table3} to the model kinds listed above.

\begin{table}
	\caption{Mapping Concerns to Model Kinds}
	\label{table:vp-table4}
	\centering 
	\footnotesize
	\begin{tabular}{p{8cm} p{3cm}} \toprule
		\textbf{\newline Concern (from \tab{vp-table3})} & \textbf{Model kind(s) that address the concern}\\ \midrule 
		Shared resources---what is shared, how is use shared & Shared Resource\newline
		Execution Time Context\newline
		Deployment\\ \midrule 
		Behavior when insufficient resource is available (run time dependencies, monitoring and measurement, interoperability context) & Shared Resource\newline 
		Interface Information\newline
		Execution Time Context\\ \midrule 
		Authentication---identity validation repository (interoperability) & Interface Information\newline
		Shared Resource\newline 
		Execution Time Context\\ \midrule 
		Authorization---remote access to system and resources (interoperability) & Interface Information\newline 
		Shared Resource\newline 
		Execution Time Context\\ \midrule 
		Encryption---algorithms and key management (interoperability) & Interface Information\newline
		Shared Resource\newline 
		Execution Time Context\\ \midrule 
		Execution time dependencies---startup sequencing (run time dependencies) & Execution Time Context\newline 
		Deployment\\ \midrule 
		Fault detection and logging---internal (monitoring and measurement) & Interface Information\newline
		Execution Time Context\\ \midrule 
		Fault recovery (interoperability context) & Execution Time Context\newline 
		Deployment\\ \midrule 
		Build time dependencies---COTS, FOSS assumptions & Code Context\\ \midrule 
		Development environment dependencies (development time dependencies, process/culture/working practices) & Code Context\newline
		Deployment\newline
		Stakeholder/Concerns\\ \midrule 
		Variabilities affecting interfaces & Interface Information\\ \midrule
		Decision model (dependencies, evolution and built-in variabilities) & Code Context\newline
		Interface Information\\ \midrule 
		Configuration dependencies among constituent systems (development time and run time dependencies) & Code Context\newline 
		Execution Time Context\newline
		Interface Information\\ \midrule 
		Perceived needs and constraints of constituent systems & Stakeholder/Concerns\\ \midrule 
		Processes, cultures, working practices between different participating organizations & Stakeholder Concerns\\ \midrule 
		Constituent system stakeholders & Stakeholder/Concerns\\ \bottomrule 
	\end{tabular}
\end{table}

\afterpage{\clearpage}

The 42010 standard permits three approaches to accommodate multiple model kinds:
\bi
\item Define multiple independent viewpoints, with each viewpoint comprising a single model kind. We rejected this approach because the models are not independent: In our case, omitting one model kind leaves a set of concerns uncovered.
\item Define a framework that comprises multiple viewpoints, with each viewpoint comprising a single model kind. The standard characterizes a framework as "establishing a common practice\dots within a particular domain of application or stakeholder community" \cite[\S4.5]{ISO42010}. We rejected this approach for two reasons: The resulting artifact was applicable beyond a single application domain, and a framework does not inherently imply that all viewpoints are used together, and so we have the model omission issue described above.
\item Define a single viewpoint that comprises multiple model kinds. We selected this approach because it treats the set of model kinds as an atomic unit.
\ei

The single viewpoint was titled "SoS Constituent System Viewpoint". The complete viewpoint definition that conforms to the ISO 42010 standard is presented in the appendix to this paper (\tion{vp-appendix} below).

\subsection{Treatment Evaluation---Active Design Review by\\Expert Panel} \label{sec:vp-treatment-eval}

Treatment evaluation is "the investigation of a treatment as applied by stakeholders in the field\ldots to investigate how implemented artifacts interact with their real-world context" \cite[\S3.1.5]{Wieringa:2014}. Our evaluation criteria were that the viewpoint provides sufficient coverage of concerns, and that a view created using the viewpoint provides sufficient detail to allow an SoS architect to reason about the constituent system operating in the context of the SoS. 

The Introduction, above, described the high stakes involved in acquiring documentation about constituent systems, arising from the managerial independence of the systems. Therefore, an initial evaluation of this untested treatment through an observational case study \cite[\S17]{Wieringa:2014} or through technical action research \cite[\S19]{Wieringa:2014} would not be a responsible approach. We chose to perform a single case mechanism experiment \cite[\S18]{Wieringa:2014} using an expert panel, to complete the initial evaluation without impacting a real-world project. Expert panel assessment has been used by a number of researchers (e.g., Dyba \cite{Dyba:2007}, van den Bosch and colleagues \cite{van-den-Bosch:2010}, and Beecham and colleagues \cite{Beecham:2005}), and so we saw this approach as both prudent and appropriate for an initial evaluation of an untested artifact. According to Hakim \cite{Hakim:1987}, small samples are effective to test explanations, particularly in the early stages of work. Expert panel recruitment is discussed below in \tion{vp-treatment-eval-s4}, and panel demographics are shown below in \tion{vp-expert-panel}.

Our treatment artifact is architecture documentation. Nord and colleagues provide a six-step structured approach to reviewing architecture documentation \cite{Nord:2009}, which we followed for our evaluation. Steps 1-5 are discussed in subsequent subsections, and the results of the review (Step 6) are discussed below in \tion{vp-results}. This six-step process is comparable to the eight-step process used by Beecham and colleagues \cite{Beecham:2005}, collapsing multiple process steps in several places.

\subsubsection{Step 1: Establish the Purpose of the Review} \label{sec:vp-treatment-eval-s1}

We aim to produce an artifact to guide an SoS architect to request the architecture information about a constituent system that is sufficient for the architect and other stakeholders to reason about that system in the context of a SoS. 

As discussed above, our criteria for the treatment evaluation were that the viewpoint provides sufficient coverage of concerns, and that a view created using the viewpoint provides sufficient detail to allow an SoS architect to reason about the constituent system operating in the context of the SoS.

\subsubsection{Step 2: Establish the Subject of the Review} \label{sec:vp-treatment-eval-s2}

We are evaluating the treatment: the artifact applied in context, specifically the viewpoint applied to a system to produce an architecture view. For this, we applied the viewpoint to the Adventure Builder system, which was chosen because it has an openly available architecture description to use as the basis for constructing the view documentation. 

The Adventure Builder system is a reference application, developed for a a fictitious company that sells adventure travel packages that bundle airline transportation, hotel accommodations, and {guid\-ed} activities. The system has a customer-facing website that allows customers to shop and purchase adventure packages, and a service-oriented architecture back-end that integrates with external payment processing and travel provider services.

The view documentation that was produced is available as part of the review instrument, shown below in \tion{instrument}.

\subsubsection{Step 3: Build or Adapt Appropriate Question Sets} \label{sec:vp-treatment-eval-s3}

Nord and colleagues identified several review styles:
\bi
\item Questionnaire---reviewers assess the artifact using a list of questions provided by the review organizer.
\item Checklist---reviewers rate the artifact using a list of yes/no questions (a special case of the questionnaire style).
\item Subjective review---stakeholders also play the role of reviewer and pose questions to themselves.
\item Active review---architects ask questions that require reviewers to use the subject artifact in order to answer the questions.
\ei
We chose to use an active review style, as this approach ensures that the reviewers skim the entire artifact and read some parts of the artifact in detail. It also evaluates the treatment (artifact in use), and not just the contents of the artifact.

However, we also wanted to understand how our experts would use their knowledge and experience to approach a SoS design problem, and so we incorporated a subjective review, where each expert formulated questions about a SoS design problem, and then later in the review, answered these as active review questions.

The instrument we created for the review had multiple parts (see \tion{instrument}, below):
\be
\item Demographic information about the reviewer.
\item A narrative vignette that created a usage scenario for the architecture view. It asked each reviewer to play the role of a SoS architect tasked to integrate the Adventure Builder system into an SoS, and specified a design problem with three new SoS capabilities. 
\item Subjective review questions---we asked each reviewer to record three questions that they had about the architecture of the Adventure Builder system, related to the design problem.
\item We next provided the architecture view artifact.
\item For each of the three new capabilities, we created several questions about the new SoS design, and the reviewer had to use the architecture view to answer the questions. 

Our questions were refined from the "Key Design Decisions" question list from Nord and colleagues \cite[\S4.5]{Nord:2009}. Two examples of the questions are shown here; the entire instrument is contained in the Appendix in \tion{instrument}.
\bi
\item Does the Adventure Builder system have existing request-response interfaces with external systems? If so, what protocols/technologies are used for these interfaces?
\item The inputs to the new payment processing interface are: Card Type, Card Number, Card Expiration, and Card Security Code/CCV. Can the current Adventure Builder system provide all of these elements?
\ei

Reviewers also answered the three subjective review questions that they created earlier in the review. We asked reviewers to annotate their responses indicating the document sections that they consulted to answer each question. We wanted to limit the duration of the review exercise to one hour, and so we presented eight active review questions, which along with the three subjective review questions, required each reviewer to answer a total of 11 questions.

\item The instrument concluded by recording the amount of time that the reviewer spent, and with open-ended questions about the realism of the scenario, the contents of the architecture view, and any comments about the review exercise.
\ee

\subsubsection{Step 4: Plan the details of the review} \label{sec:vp-treatment-eval-s4}

Nord and colleagues define three major activities in this step: Constructing the review instrument in light of constraints and intentions for reporting results; identifying actual reviewers; and planning review logistics.

Our review instrument is outlined above in \tion{vp-treatment-eval-s3}. It includes the question set, reflects how we addressed review time constraints, and collects demographic and other information to support our reporting of research results. We decided to present all reviewers with the same set of eight active review questions (i.e. all participants were assigned the same treatment).

Experts were recruited from a population of experienced practitioners in the areas of SoS and enterprise system integration, who have worked in the field for several years and have been responsible for designing the architecture for several SoS. Experts were targeted to represent different backgrounds and system domains, as recommended by Kitchenham and colleagues \cite{Kitchenham:2002}. Additionally, we sought geographic diversity and organizational diversity. We recruited eight experts from eight different organizations to participate in the review exercise, and seven accepted. 

Finally, our review logistics were simple: We emailed the instrument to each reviewer, who worked independently to complete the review and return the completed instrument to us by email.

\subsubsection{Step 5: Perform the Review} \label{sec:vp-treatment-eval-s5}

As noted above, we performed the review by sending an identical survey instrument by email to each reviewer, and receiving a completed instrument sent back to us by email. 

\section{Analysis and Results} \label{sec:vp-results}
\subsection{Expert Panel Demographics} \label{sec:vp-expert-panel}
The expert panel demographics are presented in \tab{vp-table5}. These data were self-reported by each expert. These data were collected to verify that the invitee met the inclusion standards in \tion{vp-treatment-eval-s4}, and to assess the diversity of experience of the panel.

\begin{table}[ht!]
	\caption{Expert Panel Demographic Information (N=7)}
	\label{table:vp-table5}
	\centering 
	\small
	\begin{tabular}{p{4cm} p{6.5cm}} \toprule
		\textbf{Variable} & \textbf{Reported Value(s)}\\ \midrule 
		Current position title & Software Architect (3)\newline 
		Chief Enterprise Architect (1)\newline  
		Chief Technical Officer (1)\newline 
		Solution Architect (1)\newline 
		Information Architect (1)\\ \midrule 
		Current industry & Consulting services---multiple industry domains (4)\newline 
		Government (2)\newline  
		Financial services (1)\\ \midrule 
		Prior industry experience (multiple responses allowed) & Software product development (4)\newline 
		Financial services (2)\newline 
		Academia (2)\newline 
		Consulting services---multiple industry domains (1)\newline 
		Government (1)\newline 
		Utility (1)\newline 
		Telecommunications (1)\newline  
		Transportation (1)\newline 
		Defense (1)\\ \midrule 
		Nationality &
		USA (3)\newline
		Netherlands (2)\newline
		Brazil (1)\newline
		UK (1) \\ \midrule
		Number of years of professional experience developing or integrating systems & 18-40 years (Average = 30, Median = 30)\\ \midrule 
		Approximate number of system integration projects worked on & Responses ranged from 8 to "more than 100". \\ \bottomrule 
	\end{tabular}
\end{table}

\subsection{Active Review Question Responses}
Here we discuss responses to the eight active review questions that we created and which were assigned to all reviewers.

All of the participants answered the eight active review questions correctly, with some minor variations in responses due to differing interpretations of the question wording and the context. In particular, the use of the unqualified term "interface" in several questions proved confusing: This was interpreted to mean programmatic interface or user interface, or both.

Six of the seven reviewers indicated which model(s) they used to answer each question (Some reviewers used more than one model to answer a question.). These responses showed that every model was used by at least one reviewer to answer at least one question, indicating that the questions covered the breadth of the artifact. 

Based on the small sample size (N=7), we hesitate to perform statistical analysis on the relationship between questions and models; however, we show the frequency of each model's use in \tab{vp-table6}.

\begin{table}[ht!]
	\caption{Active Review Coverage of Models}
	\label{table:vp-table6}
	\centering 
	\small
	\begin{tabular}{L{5cm} C{5cm}} \toprule
		\textbf{Model Name} & \textbf{Number of times used to answer active review question} \\ \midrule
		Stakeholder/Concerns & 13 \\ \midrule 
		Execution-time Context Model & 12\\ \midrule 
		Code Context Model & 8\\ \midrule 
		Interface Information Model & 22\\ \midrule 
		Shared Resource Model & 3\\ \bottomrule
	\end{tabular}
\end{table}

\subsection{Subjective Questions}
As discussed above in \tion{vp-treatment-eval-s3}, in addition to the eight active review questions that we developed, we asked each reviewer to specify three questions that they thought were important for this design problem. In addition to helping triangulate to improve the quality of evaluation (discussed in the next section), it provided direct insight into an SoS architect's concerns when presented with a design problem.

Our seven reviewers posed three questions each. There was significant overlap among these 21 questions, and we clustered the questions into six categories, shown in \tab{vp-table7}. 

\begin{table}[ht!]
	\caption{Subjective Question Categorization}
	\label{table:vp-table7}
	\centering 
	\small
	\begin{tabular}{L{2.75cm} L{5.75cm} C{2cm}} \toprule
		\textbf{Category} & \textbf{Example Questions} & \textbf{Frequency (N=21)}\\ \midrule
		Platform & What is the platform/technology stack/runtime environment used by the constituent system? & 7\\ \midrule 
		Data Model & What is the logical data model used in the constituent system?
		How is customer-identifying or user-identifying data handled? & 6\\ \midrule 
		Implementation Quality/Risk & What are the known problems in the constituent system?
		What is the development history (internal, acquired, outsourced, etc.)? & 3\\ \midrule 
		User Interface & What is the user interface exposed by the constituent system? & 2\\ \midrule 
		Functional Structure & What is the functional structure of the constituent system? & 2\\ \midrule 
		Architecturally-Significant Requirements & What are the quality attribute requirements for the constituent system? & 1\\ \bottomrule 
	\end{tabular}
\end{table}

The questions in the Platform category could not be readily answered by the reviewers using the architecture documentation provided. The viewpoint included a Code Context model kind, which could represent the dependencies on platform code modules, including application containers, operating systems and database libraries, and virtual machines. However, the viewpoint does not include a deployment model that would directly address this category of concerns by showing the relationship between the execution elements of the constituent system---processes, services, applications, etc.---to computer nodes and networks (e.g., \cite[\S5.2]{Clements:2011}).

The questions in the Implementation Quality/Risk category address issues that are important to understand when designing a SoS; however, this information is not part of the architecture (i.e. structures comprising elements, relationships, and properties) of the constituent system, and these concerns can be addressed by reviewing or inspecting non-architectural artifacts such as an issue tracking system or source code repository.

The questions in the User Interface category could not be readily answered by the reviewers using the architecture documentation provided. Further research is needed into the underlying concern---a possible explanation is that the enterprise business system context for the vignette that we used for the exercise triggered this concern based on the expert's experiences, even though none of the desired new capabilities involved the user interface.

The question about Architecturally-Significant Requirements was not readily answered by that reviewer using the architecture documentation provided. A complete ISO/IEC/IEEE 42010-compliant architecture description would contain rationale that includes architecturally significant requirements \cite[\S5.8]{ISO42010}. Our review instrument included only a subset of a complete architecture description containing the view that was the subject of the evaluation, and the view contained models with no rationale.

Questions in the Data Model and Functional Structure categories were readily answered by the reviewers using the architecture documentation provided.

\subsection{Interpretation and Viewpoint Rework} \label{sec:vp-results-rework}
Based on the experiment results discussed above, we found that the baseline version of the architecture viewpoint adequately covered the SoS stakeholder concerns that we had identified in our Problem Investigation. However, our expert review panel's subjective review questions uncovered three categories of concerns that we had not identified in our Problem Investigation, and were not addressed by the baseline version of architecture viewpoint.

Below, we discuss each of these categories of concerns, and how the baseline version of viewpoint definition was reworked to produce the fubak viewpoint definition presented in the Appendix (\tion{vp-appendix}).

\subsubsection{Runtime Deployment Environment Concerns}
The first category of concerns that was not addressed by the baseline version of the architecture viewpoint involved the runtime deployment environment of the constituent system. The subjective review questions that raised this concern covered two areas: the software platform (operating system, application server, database manager, service bus, etc.), and the physical deployment (mapping of software to compute nodes and networks). 

In developing the viewpoint definition, we expected that the software platform concerns would be addressed by the Execution Time Context Model and/or the Code Context Model; however, those models in the Adventure Builder System artifact provided to the reviewers did not contain sufficient detail to address the concern. We have reworked the viewpoint definition by extending the "Elements" section of these two models to add the software platform elements to the model, and by extending the "What's it for" section to add that the model is used to answer questions such as those posed by the expert panel.

The physical deployment concerns arise from the distributed nature of a SoS. This physical distribution affects performance, availability, and possibly security and other qualities. It is necessary for the SoS architect to understand the physical deployment of the constituent system (how software is mapped to compute and network resources) because, when the system becomes part of an SoS, those compute and network resources may be shared with other constituent systems, or may be configured differently from the constituent system's stand-alone architecture. 

The Shared Resource Model definition identifies network bandwidth and compute resources as elements that may be shared, in order to address concerns about performance. In the Adventure Builder System artifact provided to the reviewers, that model was represented as a table, with deployment information provided as part of the description of each element.  This presentation style did not provide sufficient detail to address the reviewer's concern. Many architecture documentation approaches define a Deployment Model, for example the Deployment Style defined by Clements and colleagues \cite{Clements:2011} or the Physical View defined by Kruchten \cite{Kruchten:2003}. In this model, elements are units of software execution (e.g., processes, services, etc.), and physical infrastructure (e.g., computer nodes and networks), and the relation of "executes on" maps software to physical infrastructure. This model is used to address concerns about performance, availability, and possibly security and other qualities. We have reworked the viewpoint definition to add a Deployment Model.

\subsubsection{Implementation Quality and Risk Concerns}
The second category of concerns that was not addressed by the baseline version of the architecture viewpoint involved the quality of the implementation of the constituent system, and assessing risk in integrating it into the SoS.

These concerns might be seen to intersect with several of the concerns drawn from the practitioner-oriented literature shown in \tab{vp-table2}, namely "Processes, cultures, working practices" and "Constituent System Evolution Strategy", but on the whole, they were not considered in developing the baseline version of the viewpoint. 

As discussed above, it is important to understand these issues when designing a SoS. The expert panel's questions reflect common architecture approaches, such as the Risk and Cost Driven Architecture approach \cite{Poort:2012}. However, we think that this is not part of the architecture (i.e. structures comprising elements, relationships, and properties) of the constituent system, and that these concerns can be addressed by reviewing artifacts such as an issue tracking system or source code repository. We did not rework the baseline version of the viewpoint in response to this gap.

\subsubsection{User Interface Concerns}
The third category of concerns that was not addressed by the baseline version of the architecture viewpoint involved the user interface of the constituent system. 
As noted above, the constituent systems in a SoS are characterized by operational independence, which would imply one of two user interface integration patterns:
\bi
\item \emph{Mashup}, where the SoS user interface is developed using APIs of the constituent systems. In this case, the constituent system's user interface is not presented directly.
\item \emph{Portal}, where the user interface of a constituent system is presented in a sub-window (e.g., frame or pane) of the SoS user interface. In this case, the constituent system's user interface is presented in its entirety, without modification.
\ei

Therefore, concerns about the user interface, per se, do not appear to be generalizable SoS concerns, but there may be system-specific concerns. For example, a mashup approach would introduce concerns that would be addressed by the Execution Time Context Model and the Interface Information Model. A portal could introduce concerns about the user interface display as a shared resource, to be addressed by the Shared Resource Model.

We did not rework the baseline version of the viewpoint in response to this gap; however, this may be an area for further research.

\subsection{Threats to Validity}
Construct validity is the degree to which the case is relevant with respect to the evaluation objectives \cite{Runeson:2008}. Here, our objective was to evaluate the ability of an architecture documentation viewpoint to address the concerns of a SoS architect about a constituent system within the SoS, in order to support SoS design and analysis involving that constituent system. The selection of the active review evaluation method ensured that the reviewers at least skimmed the entire document, and our recording of the sections of the document used to answer each question ensured that certain sections were read in detail. Also, the reviewer's positive comments about the realism of the review vignette support the construct validity of the experiment. Our objective \emph{was not} to compare this treatment (use of the viewpoint to reason about the constituent system) to other treatments (e.g., using documentation and code from the constituent system to reason about the system).

Internal validity concerns hidden factors, which is a concern when examining causal relations \cite{Runeson:2008}. Our use of the active review method introduced the potential threat to internal validity that the questions created for the review may have been unconsciously influenced by our knowledge of the Adventure Builder system architecture and its documentation. We mitigated this risk by also incorporating subjective review questions: prior to reading the architecture documentation, each reviewer created three questions, and then later used the documentation to answer those questions. This use of triangulation increases the reliability of our results \cite{Runeson:2008}.

External validity is related to the generalizability of the results in the context of a specific population. As discussed above in \tion{vp-treatment-eval}, we chose to use an expert panel with a small sample size. According to Hakim \cite{Hakim:1987}, small samples are effective to test explanations, particularly in the early stages of work. By definition, our single case mechanism experiment does not support statistical generalization, and so suffers the same external validity challenges of all case study research \cite{Yin:2003}. Our total response rate (recruitment to completion) was 87.5\%, from an expert panel with a diversity of experience and system domain coverage, so we believe that our findings are valid at least for SoS and constituent systems that are similar in size and scope to the SoS described in the vignette that formed the basis of our review. 

\section{Conclusions and Future Work} \label{sec:vp-conclusions}
In this paper, we have introduced an architecture viewpoint to address the concerns of a SoS stakeholders about a constituent system within the SoS, in order to support SoS design and analysis involving that constituent system. We evaluated this viewpoint using a single case mechanism experiment: An expert panel performed an active design review using a question set that we provided. The expert panel also created subjective questions, which provided additional insight into the concerns of a SoS architect when solving a design problem and improved the quality of our data by mitigating internal validity concerns inherent in the active review process.

The evaluation results were generally positive, with the viewpoint showing promise in providing guidance for SoS architects seeking architecture knowledge about a constituent system. However, the evaluation identified a gap in the baseline version of the viewpoint definition: It was missing a deployment model for the constituent system that shows the relationship of the software to computer nodes and networks. The viewpoint definition presented in the appendix has been reworked to reflect this change. 

The viewpoint conforms to the ISO/IEC/IEEE 42010:2011 (E) standard for architecture description, and the revised viewpoint comprises five model kinds: Constituent System Stakeholders/Concerns, Constituent System Execution Time Context, Constituent System Code Context, Constituent System Interface Information Model, Shared Resource Model, and Deployment Model. 

The managerial independence of constituent systems poses challenges for SoS architecture designers, and frequently the architecture knowledge acquisition process involves high stakes activities that risk damage to the architect's reputation and other consequences. Further empirical research in this area must be designed within the constraints of this context. Our results provide the confidence to evaluate this viewpoint using methods such as case study or technical action research.

\section{Appendix: Viewpoint Definition} \label{sec:vp-appendix}
The viewpoint defined here is a revised version of the baseline viewpoint used to create the artifact that was the subject of the experiment discussed in the body of this paper. The following revisions were made to the baseline version of the viewpoint, as described in \tion{vp-results-rework}:
\bi
\item The "Elements" sections of the Execution Time Context Metamodel (\tab{vp-table10}) and the Code Context Metamodel (\tab{vp-table11}) were revised to specify that platform elements such as operating system, application server, and database manager should be included.
\item A new metamodel was added. The Deployment Metamodel (\tab{vp-table14}) relates software units of execution (e.g., processes or services) to the execution environment of computers and networks. \tab{vp-table15} and \tab{vp-table16} were revised to add a reference to the new Deployment Metamodel.
\ei

This viewpoint definition follows the template in Annex B of ISO 42010 \cite{ISO42010}.

\subsection{Viewpoint Name}

This defines the "SoS Constituent System Viewpoint", for use in documenting the relevant parts of the architecture of one constituent system in a SoS.
\subsection{Viewpoint Overview}
The need for this viewpoint is discussed in \tion{vp-intro} of this paper.

\subsection{Concerns Addressed by this Viewpoint}
\tab{vp-table2} and \tab{vp-table3} in the body of this paper show the concerns addressed by this viewpoint, and map the concerns to the stakeholder roles identified in the next section, below. \tion{vp-treatment-eval-s1} also discusses the method used to identify the concerns.

\subsection{Typical Stakeholders}
The stakeholder roles addressed by this viewpoint are shown in \tab{vp-table1} in the body of this paper.

These stakeholders were selected because they are directly involved in understanding the constituent system architectures, proposing or defining changes to those architectures for use in the SoS, and then constructing, testing, and integrating the changed constituent systems in the SoS.

\subsection{Model Kinds/Metamodels}
This viewpoint specifies of a number of model kinds\footnote{In the terminology of ISO 42010, a \emph{viewpoint} applied to a system yields a \emph{view}. Analogously, the standard defines a \emph{model kind}, which, when applied to a system, yields a \emph{model}.}. 

We apply the principle of separation of concerns, and so each model kind is defined using a single architecture style \cite{Clements:2011}: module styles address development time concerns, component and connector styles address execution time concerns, and allocation styles map between software elements and their environment. 

Each model kind is specified as a metamodel. The metamodel template is shown in \tab{vp-table8}, and is based on the Style Guide Template defined by Clements and colleagues \cite{Clements:2011}. 

\begin{table}[htb!]
	\caption{Template used to specify metamodels for model kinds in this viewpoint}
	\label{table:vp-table8}
	\centering 
	\small
	\begin{tabular}{L{2cm} p{9cm}} \toprule
		\textbf{Name:} & \textbf{Name of the model kind}\\ \midrule 
		Type: & Module, component and connector, or allocation, as defined by Clements and colleagues \cite{Clements:2011}.\\ \midrule 
		Elements: & The types of elements allowed in this model kind, and the properties that should be attached to each element instance.\\ \midrule 
		Relations: & The types of relations among elements allowed in this model kind, and the properties that should be attached to each relation instance.\\ \midrule 
		Constraints: & Any model construction constraints, such as cardinality of element or relation types or topology constraints.\\ \midrule 
		What's it for: & Brief description of how the model kind is used to support SoS architecture tasks such as design, analysis, evolution, or evaluation.\\ \midrule 
		Notations: & Recommended notations for documenting the model kind, such as table, diagram, or list.\\ \bottomrule 
	\end{tabular}
\end{table}

The first model kind is defined in \tab{vp-table9}, and represents the stakeholders and their concerns for the constituent system. This provides architecture context for the SoS architect and Program Manager by providing insight into the perceived need of the constituent system, and identifies stakeholders who may be impacted by the constituent system's operation within the SoS.

\begin{table}[htb!]
	\caption{Constituent System Stakeholders/Concerns Metamodel}
	\label{table:vp-table9}
	\centering 
	\small
	\begin{tabular}{p{2cm} p{9cm}} \toprule
		\textbf{Name:} & \textbf{Constituent System Stakeholders/Concerns}\\ \midrule 
		Type: & Allocation\\ \midrule 
		Elements: & Constituent system stakeholders\newline 
		Stakeholder concerns about system architecture\\ \midrule 
		Relations: & A stakeholder has a concern\\ \midrule 
		Constraints: & Stakeholders can have multiple concerns.\newline 
		Multiple stakeholders can have the same concern.\\ \midrule 
		What's it for: & Aids in understanding the scope of the constituent system, and who will be impacted by changes made to the constituent system to allow it to join the SoS.\\ \midrule 
		Adding \newline Assumptions: & List any stakeholders that were considered but intentionally excluded.\newline
		Note concerns that were identified but not addressed by the architecture.\\ \midrule 
		Notations: & List---one item per stakeholder, with list of concerns.\newline
		Matrix---one row per stakeholder, one column per unique concern, "x" at row-column intersection means that the stakeholder in that row has the concern in that column\\ \bottomrule 
	\end{tabular}
\end{table}

The second metamodel in this viewpoint, shown in \tab{vp-table10}, addresses concerns related to dependencies at execution time, shared resources, and to a lesser extent, overall context.

\begin{table}[htb!]
	\caption{Constituent System Execution Time Context Metamodel}
	\label{table:vp-table10}
	\centering 
	\small
	\begin{tabular}{p{2cm} p{9cm}} \toprule
		\textbf{Name:} & \textbf{Constituent System Execution Time Context}\\ \midrule 
		Type: & Component and Connector\\ \midrule 
		Elements: & Running system\newline
		External software that the system interacts with\\ \midrule 
		Relations: & Any interaction at execution time (e.g., sends/receives message, call/return, reads/writes data, interrupts, synchronizes with)\newline
		Property: Interfaces used for the interaction on self and external software\newline
		Property: Direction of interaction (initiated by constituent system or external system)\\ \midrule 
		Constraints: & An interface on the constituent system may be used to interact with multiple external systems\newline
		Multiple external systems may interact with the constituent system through the same interface on the constituent system\\ \midrule 
		What's it for: & Aids in understanding the scope of the constituent system to analyze the impacts of necessary or desired changes \newline
		Identifying viable SoS subsets and activity sequencing during SoS integration\\ \midrule 
		Adding \newline Assumptions: & Startup behavior should be documented, using a notation such as a message sequence diagram\newline
		Monitoring and performance measurement behavior should be documented, using notations such as message sequence diagrams and state transition diagrams.\\ \midrule 
		Notations: & Diagram---e.g., Context Diagram from Clements \cite{Clements:2011}\newline
		List---one item per constituent system interface, with list of external systems and interfaces that it interacts with\newline
		Matrix---rows are interfaces on the constituent system, columns are interfaces on external systems, "S" at a row-column intersection means that the constituent system interface sends an interaction to the external system, "R" means that the constituent system interface receives an interaction from the external system\\ \bottomrule 
	\end{tabular}
\end{table}

Concerns related to development time are addressed in the metamodel defined in \tab{vp-table11}.

\begin{table}[htb!]
	\caption{Constituent System Code Context Metamodel}
	\label{table:vp-table11}
	\centering 
	\small
	\begin{tabular}{p{2cm} p{9cm}} \toprule
		\textbf{Name:} & \textbf{Constituent System Code Context}\\ \midrule 
		Type: & Module\\ \midrule 
		Elements: & Constituent system software\newline
		External modules (libraries, packages, development tools, etc.) that the constituent software depends on\\ \midrule 
		Relations: & Uses\newline
		Properties: type of dependency (e.g., code generation, build, unit test, integration test), version identification or key features used for external modules, source of external modules (e.g., FOSS, COTS, GOTS)\\ \midrule 
		Constraints: & Many-to-many\\ \midrule 
		What's it for: & Aids in understanding the scope of the constituent system to analyze the impacts of necessary or desired changes. \newline Identifying mismatches among external dependencies that will constrain deployment decisions or interactions among constituent systems in the SoS.\\ \midrule 
		Adding \newline Assumptions: & What evolution is assumed for the external modules? Are there new features or capabilities that are expected to be available that the constituent system will use?\\ \midrule 
		Notations: & Diagram---e.g., Uses Context Diagram from Clements \cite{Clements:2011}\newline 
		List\newline
		Matrix---This structure may be documented in a Dependency Structure Matrix generated for static analysis of the constituent system code. \\ \bottomrule 
	\end{tabular}
\end{table}

The metamodel defined in \tab{vp-table12} addresses general information interoperation concerns.

\begin{table}[htb!]
	\caption{Constituent System Interface Information Metamodel}
	\label{table:vp-table12}
	\centering 
	\small
	\begin{tabular}{p{2cm} p{9cm}} \toprule
		\textbf{Name:} & \textbf{Constituent System Interface Information Model}\\ \midrule 
		Type: & Module\\ \midrule 
		Elements: & Information elements of interest to the SoS (e.g., a SoS that deals with geo-location might have concepts like position, elevation, and direction)\newline
		Information elements in the constituent system software architecture\newline
		Properties: Should include units, timeliness, precision, security level, etc., as applicable\\ \midrule 
		Relations: & Between SoS and constituent system information elements, and from constituent system elements to sub-elements (to refine details).\newline
		Logical associations (1-1, 1-n, n-m)\newline
		Specialization/generalization (is-a)\newline
		Aggregation\\ \midrule 
		Constraints: & None\\ \midrule 
		What's it for: & Understanding how common concepts in the SoS are represented in a constituent system, and identifying mismatch between representations among constituent systems in the SoS \\ \midrule 
		Adding \newline Assumptions: & Explicitly identify SoS information elements that have no relationship to the constituent system \\ \midrule 
		Notations: & Logical data modeling notations (ERD, UML)\\ \bottomrule 
	\end{tabular}
\end{table}

Concerns about resource sharing are addressed in the metamodel defined in \tab{vp-table13}.

\begin{table}[htb!]
	\caption{Shared Resource Metamodel}
	\label{table:vp-table13}
	\centering 
	\small
	\begin{tabular}{p{2cm} p{9cm}} \toprule
		\textbf{Name:} & \textbf{Shared Resource}\\ \midrule 
		Type: & Component and Connector\\ \midrule 
		Elements: & Component(s) representing a resource that is used by the constituent system and by other external systems. These include processor computing cycles, memory, disk space, network interfaces, network bandwidth, files, databases or repository, virtual infrastructure, and system physical resources such as a display, radio frequency, or antenna.\newline 
		Component(s) in the constituent system that use the shared resource\\ \midrule 
		Relations: & Any interaction during execution that acquires, consumes, or releases the shared resource.\\ \midrule 
		Constraints: & None\\ \midrule 
		What's it for: & Analyzing capacity and performance/availability of the SoS. Identifying cases of undesirable SoS behavior due to mismatch between resource sharing approaches of constituent systems.\\ \midrule 
		Adding \newline Assumptions: & Is the resource explicitly or implicitly acquired and released? \newline 
		What is the behavior if insufficient (or no) resources are available?\\ \midrule 
		Notations: & Static diagrams---e.g., from Views and Beyond component and connector style guide \cite{Clements:2011}\newline 
		Behavior diagrams---message sequence charts, state transition diagrams, etc.\\ \bottomrule 
	\end{tabular}
\end{table}

Concerns about deployment of software onto computers and networks are addressed in the metamodel defined in \tab{vp-table14}.

\begin{table}[htb!]
	\caption{Deployment Metamodel}
	\label{table:vp-table14}
	\centering 
	\small
	\begin{tabular}{p{2cm} p{9cm}} \toprule
		\textbf{Name:} & \textbf{Deployment}\\ \midrule 
		Type: & Allocation\\ \midrule 
		Elements: & Software units of execution, with properties that specify the execution needs and constraints\newline
		Computers and networks that execute software, with properties that specify the execution resources (e.g., compute cycles, memory, storage, and bandwidth) provided\\ \midrule 
		Relations: & Software units execute on computers and networks.\\ \midrule 
		Constraints: & None\\ \midrule 
		What's it for: & Analyzing performance and availability, and possibly security and other qualities. Assessing operating cost (e.g., required hardware and software, operations staff skills).\\ \midrule 
		Adding \newline Assumptions: & Can any part of the execution environment be virtualized?\newline
		What are the assumptions about the network's ability to reach the internet or particular resources on a private network?\\ \midrule 
		Notations: & Table---Rows are instances or types of software elements, columns are instances or types of computers and networks\newline 
		Diagram---e.g., Deployment Diagram from Clements \cite{Clements:2011}\\ \bottomrule 
	\end{tabular}
\end{table}

\subsection{Correspondence rules}
There are no specific correspondence rules for the models constructed using this viewpoint.

\subsection{Operations on views}
\subsubsection{Creating a view of a constituent system using this viewpoint}
In some cases, the information needed to create a view using this viewpoint already exists in the architecture documentation for the constituent system. \tab{vp-table15} and \tab{vp-table16} map the information required for this viewpoint to sources in two commonly used documentation frameworks: Views and Beyond \cite{Clements:2011}, and DoDAF \cite{DOD:2010}.

\begin{table}[htb!]
	\caption{Mapping from SoS Viewpoint Metamodels to Views and Beyond Approach}
	\label{table:vp-table15}
	\centering 
	\small
	\begin{tabular}{p{3cm} p{8cm}} \toprule
		\textbf{Viewpoint Metamodel Name} & \textbf{Source of information in a Views and Beyond architecture document}\\ \midrule 
		Constituent System Stakeholders/Concerns & Information Beyond Views---Documentation Roadmap.\newline 
		Stakeholder/View Matrix (typically generated by the architect but not explicitly included in the architecture documentation).\\ \midrule
		Constituent System Execution Time Context & Context diagram from one of the component and connector views, e.g., client-server, SOA, pipe and filter, or publish-subscribe.\\ \midrule 
		Constituent System Code Context & Context diagram from a module uses view.\\ \midrule 
		Constituent System Interface Information Model & Interface documentation for externally-visible interfaces (from component and connector views), or a data model view packet focused on externally-visible information elements.\\ \midrule 
		Shared Resource & Component and connector view.\\ \midrule 
		Deployment & Deployment view primary presentation or context diagram.\\ \bottomrule 
	\end{tabular}
\end{table}

\begin{table}[htb!]
	\caption{Mapping from SoS Viewpoint Metamodels to DoDAF}
	\label{table:vp-table16}
	\centering 
	\footnotesize
	\begin{tabular}{>{\raggedright}p{2.75cm} >{\raggedright}p{3cm} >{\raggedright\arraybackslash}p{4.75cm} } \toprule
		\textbf{Viewpoint Model Name} & \textbf{Source of information in a DoDAF architecture document} & \textbf{Comments}\\ \midrule 
		Constituent System Stakeholders/Concerns & AV-1 Overview and Summary Information\newline
		PV-1 Project Portfolio Relationships & DoDAF does not generally treat stakeholders as a first-order concern. These DoDAF views provide insight into the operational, maintenance, and development stakeholders.\\ \midrule 
		Constituent System Execution Time Context & SvcV-1 Services Context Description\newline
		SvcV-3b Services-Services Matrix & \\ \midrule 
		Constituent System Code Context & SvcV-1: Services Context Description & If the information is included, it is most likely to appear in the SvcV-1.\\ \midrule 
		Constituent System Interface Information Model & SvcV-2: Services Resource Flow Description\newline 
		SvcV-6: Services Resource Flow Matrix\newline 
		StdV-1 Standards Profile & DoDAF use the concept of "resource flows" to identify interfaces and protocols.\\ \midrule 
		Shared Resource & SvcV-3b Services-Services Matrix\newline
		SvcV-10c Services Event-Trace Description & Shared resources may not be explicitly identified, but can be discovered using the SvcV views.\\ \midrule 
		Deployment & SvcV-1 Services Context Description\newline
		SvcV-3a Systems-Services Matrix & DoDAF "services" usually include both software and hardware elements, without explicit refinement. The DoDAF views noted here may provide insight, but are unlikely to provide all the information needed to create this model.\\ \bottomrule 
	\end{tabular}
\end{table}

\subsubsection{Interpretive, Analysis, and Design Methods}
These operations on a view created from this viewpoint are discussed in the "What's it for" section of the metamodels specified above.

\subsection{Examples and Notes}
The evaluation instrument in \tion{instrument} provides an example of applying this viewpoint to create a view on a constituent system.

\clearpage

\section{Appendix: Evaluation Instrument (including example use of the viewpoint)} \label{sec:instrument}
This appendix contains the instrument used to evaluate the viewpoint, as discussed above in \tion{vp-treatment-eval-s3}.

In order to construct the evaluation instrument, we had to use the viewpoint to construct system architecture documentation for the Adventure Builder System. The models were created in accordance with the viewpoint definition specified in \tion{vp-appendix}, and so this also provides an example of the use of the viewpoint.

The evaluation instrument begins on the next page.

\clearpage
This is a role-playing exercise. You will play the role of an architect at a fictitious company called "Social Travel". Your business sells travel packages, and provides a number of social networking capabilities to allow your users to connect and share information with each other.
You are responsible for a collection of integrated enterprise systems:
\begin{itemize}
	\item \textit{TravelPhotos}:provides photo storage, tagging, and sharing with other Social Travel users.
	\item \textit{TravelStuff}: an online retailer for travel-related gear
	\item \textit{TravelIns}: marketing travel insurance, such as trip cancellation coverage, international medical coverage, and emergency rescue/evacuation coverage.
	\item Enterprise-wide Social Features: a back-end system for cross-cutting features such as Profiles, Friends, Tags, Recommendations, Sharing, etc.
\end{itemize}

Your company has just acquired a smaller company called "Adventure Builder" that specialized in selling adventure travel packages. You need to integrate some of their systems with your enterprise systems, to realize three new capabilities:

\noindent\fbox{\begin{minipage}{\linewidth}

\textbf{New Capability \#1---Social features}

This capability adds social features to Adventure Builder catalog browsing. A user can see which trips her friends have taken, see comments about trips from other users, see trip photos from other users, share plans and itineraries with friends, etc.

The architecture approach to achieve this capability will store the social data repository (i.e. user profile, tags, sharing links, comments, etc.) outside of the Adventure Builder system. The Adventure Builder system must provide keys (or IDs) for data inside Adventure Builder that the social repository can link to. We also need to insert new widgets and elements into the customer UI to allow access to the social features.

\textbf{New Capability \#2---Common payment processing interface}

This capability changes the Adventure Builder payment processing interface to use the same interface as is used by the existing TravelStuff systems.

The architecture approach to achieve this capability is to completely replace Adventure Builder's "Bank" interface.

\textbf{New Capability \#3---Add cross-sell features}
This capability changes the Adventure Builder user interface to cross-sell products from TravelStuff and TravelIns when user books trip. For example, based on the destination, offer relevant clothing products from TravelStuff and insurance from TravelIns against a hurricane forcing the trip to be cancelled.

The architecture approach to achieve this capability is to make a request to existing SocialTravel cross-sell business logic trip with itinerary information, and then insert new widgets and elements into the customer UI to display the cross-sell offers.
\end{minipage}}

In order to scope the work to develop these new capabilities, you need information about the Adventure Builder system. Below, please list three questions that you have about the Adventure Builder system architecture. These questions do not have to be the highest priority, or listed in any particular order. We are trying to understand how you approach this problem.
\begin{table}[htb!]
	\centering 
	\small
	\begin{tabular}{p{2cm} p{9cm}} \toprule
		 & \textbf{Your Questions:}\\ \midrule 
		Question 1: & \\ \midrule 
		Question 2: & \\ \midrule 
		Question 3: & \\ \bottomrule 
	\end{tabular}
\end{table}

\clearpage
As often happens in a corporate acquisition, you do not have direct access to your counterpart architect at Adventure Builder. In this case, as soon as the deal closed, the Adventure Builder architect cashed in her stock and left on a trip around the world. 

However, before she left, she prepared a documentation package for you that addresses concerns related to integrating an existing system into an enterprise or into a system of systems (SoS). This documentation refers to the as-is existing system as a "constituent system" of the SoS.

This documentation has five models, summarized here:
\medskip

\noindent\textbf{Constituent System Stakeholders/Concerns Model:} Maps stakeholders in the constituent system to their concerns about that system. This helps you understand the scope of the constituent system, architecture drivers, and who will be impacted by changes that you make to the constituent system to allow it to join the SoS.

\noindent\textbf{Constituent System Execution Time Context Model:} Describes any runtime interactions between the constituent system and external systems, including request/response, data exchange, message passing, and error/exception handling.

\noindent\textbf{Constituent System Code Context Model:} Identifies external modules (libraries, packages, development tools, etc.) that the constituent software depends on, along with the type of dependency.

\noindent\textbf{Constituent System Interface Information Model:} Information elements within the constituent system that are of interest to the SoS (e.g., a SoS that deals with geo-location might have concepts like position, elevation, and direction). Note that this data may not be accessible through existing external interfaces of the constituent system.

\noindent\textbf{Shared Resource Model:} Component(s) that represent a resource used by the constituent system and by other external systems. These resources include processor computing cycles, memory, disk space, network interfaces, network bandwidth, files, databases or repository, virtual infrastructure, and system physical resources such as a display, radio link, or sensor. This model also includes component(s) in the constituent system that use each shared resource.

\medskip

The models in the documentation package are shown on the following pages. There is a lot of detail here---skim it for now, and some later questions will ask you to look at it more closely. Again, you can print this document if you want to.

Some of the models have traceability references back to the Adventure Build architecture description, which is available at \url{https://wiki.sei.cmu.edu/sad/index.php/The_Adventure_Builder_SAD}, but you should not need to refer to that directly.
\clearpage

\subsection*{Model: Constituent System Stakeholders/Concerns}
\begin{center}
	\captionof{table}{Constituent System Stakeholders/Concerns Model for the Adventure Builder System}
	\small
	\begin{tabular}{L{2.75cm} L{5.75cm} L{2cm}} \toprule
	\textbf{Stakeholder}&\textbf{Concerns}&\textbf{Source}  \\ \midrule
	Product Manager&Modifiability---add new business partners quickly&QAS1  \\ \midrule
	Product Manager&Usability, Performance---purchase action latency&QAS2  \\ \midrule
	Product Manager\newline Operations&Latency under load&QAS3  \\ \midrule
	Operation\newline Finance&Reliability---Purchase requests to OPC are idempotent.\newline Usability---Successful purchases are always acknowledged to customer.&QAS4  \\ \midrule
	Legal\newline Info. Sec.\newline Product Manager&Security---Payment processing transactions are secure and meet all internal policies and regulatory compliance requirements.&QAS5  \\ \midrule
	Operations\newline Info. Sec.\newline Product Manager&Denial of Service attack is detected and handled.&QAS6  \\ \midrule
	Operation&24/7 availability.\newline Failures detected and notification issued within 30 seconds.&QAS7  \\ \bottomrule
	\end{tabular} 
\end{center}

\clearpage

\subsection*{Model: Constituent System Execution Time Context}
\begin{center}
	\captionof{table}{Constituent System Execution-Time Context Model for the Adventure Builder System}
	\small
	\begin{tabular}{L{2cm} L{5.75cm} L{2.75cm}} \toprule
		\textbf{External System}&\textbf{Interfaces to the external system and interface properties}&\textbf{Source}  \\ \midrule
		Bank&Interface: CreditCard Service/SOAP/Adventure Builder invokes request&Top Level SOA\newline View Primary Presentation \\ \midrule
		Airline Provider&Interface: AirlinePO Service/SOAP/Adventure Builder invokes request \medskip
		                 Interface: Web Service Broker/SOAP/Adventure Builder receives request &Top Level SOA\newline View Primary Presentation \\ \midrule
		Lodging Provider&Interface: LodgingPO Service/SOAP/Adventure Builder invokes request \medskip
                         Interface: Web Service Broker/SOAP/Adventure Builder receives request &Top Level SOA\newline View Primary Presentation \\ \midrule
		Activity Provider&Interface: ActivityPO Service/SOAP/Adventure Builder invokes request\medskip
                          Interface: Web Service Broker /SOAP/ Adventure Builder receives request &Top Level SOA\newline View Primary Presentation \\ \midrule
		User's Email Client&Interface: SMTP/SMTP/external configuration file&Top Level SOA\newline View Primary Presentation \\ \bottomrule
	\end{tabular} 
\end{center}

Also, the top-level workflow diagram is applicable here to show how these interfaces are used in practice. This is shown in \fig{vp-workflow}.

\begin{center}
	\includegraphics[width=.9\textwidth]{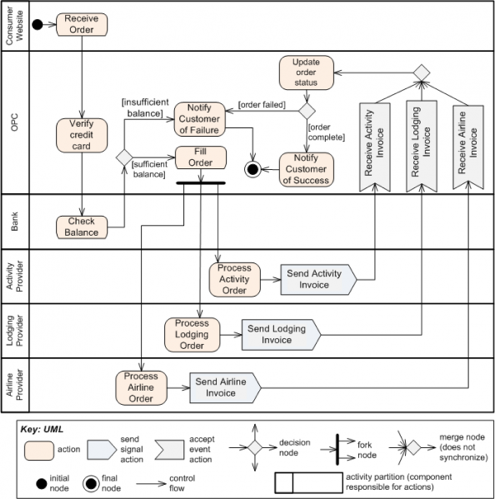}
	\captionof{figure}{Workflow behavior diagram from Adventure Builder Architecture}
	\label{fig:vp-workflow}
\end{center}

\clearpage

\subsection*{Model: Constituent System Code Context}
\begin{center}
	\captionof{table}{Constituent System Code Context Model for the Adventure Builder System}
	\small
	\begin{tabular}{L{2.75cm} L{5.75cm} L{2cm}} \toprule
		\textbf{External Module used by AB System}&\textbf{Properties}&\textbf{Source}  \\ \midrule
		gwt (Google Web Toolkit) & Dependency Type: Build (generates Javascript for execution); \newline Version unspecified; \newline Open Source
		 & Top Level Module Uses Diagram\\ \midrule
		waf (Web Application Framework) & Dependency Type: Build?; \newline Version "Java Blueprints"; \newline Open Source
& Top Level Module Uses Diagram\\ \midrule
		wsdls & Dependency: Build; \newline Version unspecified; \newline license unspecified & Top Level Module Uses Diagram \\ \bottomrule
	\end{tabular} 
\end{center}

\clearpage

\subsection*{Model: Constituent System Interface Information Model}
\begin{center}
	\small
	\includegraphics[width=.9\textwidth]{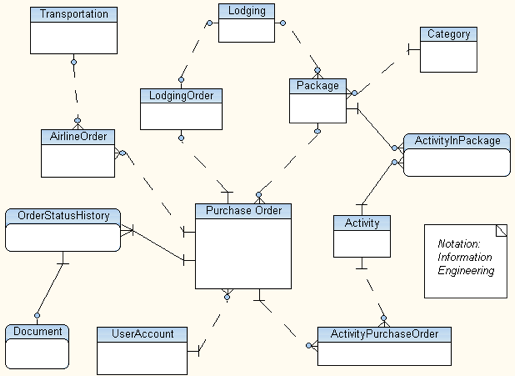}
	\captionof{figure}{Part 1 of Constituent System Interface Information Model for the Adventure Builder System (From Adventure Builder Data Model View)}
	\label{fig:vp-intf-info}
\end{center}

\clearpage

Our approach for new capability \#2 requires us to replace the Bank interface. Part of the information model for the Bank interface is shown in \fig{vp-intf-ccard}.

\begin{center}
	\captionof{table}{Element Catalog for Part 1 of Interface Information Model}
	\small
	\begin{tabular}{L{2.75cm} L{8cm}} \toprule
		\textbf{Information Element}&\textbf{Description} \\ \midrule
		PurchaseOrder
& Aggregate of transportation, lodging, package, and activity orders.
\\ \midrule
		UserAccount	& An end user of the AdventureBuilder application. We store email id, password, and contact info.
\\ \midrule
		AirlineOrder & Aggregate of purchased transportation entries.
\\ \midrule
		Transportation & Each transportation entry is a flight available for booking in our travel packages. For each one, we record: name, departure and arrival airports, days and times, airline name, flight number, rate, cabin class.
\\ \midrule
		LodgingOrder & Aggregate of purchased lodging entries.
\\ \midrule
		Lodging & A hotel, guesthouse or B\&B that can be used for lodging in our travel packages. For each type of lodging, we store name, description, location info, room description, rates.
\\ \midrule
		Package	& A travel package available in our catalog. A package specifies lodging and a list of activities. Attributes of a package include name, description, rate per person, category and a representative image to show to the user.
\\ \midrule
		Category & A category of adventure travel packages. Examples: island packages, mountain adventures. This categorization helps the user to browse through our catalog of packages. Category data consists of a name, description and a representative image to show on the user screen.
\\ \midrule
		Activity & An adventure activity available. Examples: snorkeling, fishing, bird watching, rafting, surfing. Activities are available in selected packages. Information stored for each activity include: name, description, rate, and a representative image to show to the user.
\\ \midrule
		ActivityInPackage & This entity represents the many-to-many relationship between activity and package. It simply lists the activities in each package ("join table")
\\ \midrule
		ActivityPurchaseOrder & Aggregate of purchased activity entries ("join table") \\ \bottomrule
	\end{tabular} 
\end{center}

\clearpage

	\texttt{\textbf{Type CreditCard} \newline
		// This type is used to store the credit card information of the user.
\newline
		String cardExpiryDate 
\newline
		String cardNumber 
\newline
		String cardType 
\newline
	}
\begin{center}
	\captionof{figure}{Part 2 of Constituent System Interface Information Model for the Adventure Builder System}
	\label{fig:vp-intf-ccard}
\end{center}

\clearpage

\subsection*{Model: Shared Resource Model}

In the stand-alone Adventure Builder system, there are no resources shared with any external systems.

In the new SoS, we have several resources that now will be shared. These are shown in \tab{vp-shared-res}. For each resource identified in \tab{vp-shared-res}, the source of the information in the Adventure Builder Architecture Documentation is shown.

\begin{center}
	\captionof{table}{Shared Resource Model for the Adventure Builder System}
	\label{table:vp-shared-res}
	\small
	\begin{tabular}{L{2cm} L{3cm} L{3cm} L{2cm}} \toprule
		\textbf{External Resource}&\textbf{Adventure Builder Resource Usage}&\textbf{Resource shared with}&\textbf{Source}  \\ \midrule
		Bank Interface (accessed through Firewall) & Validate credit card for every customer purchase (Call/Return) & Other SocialTravel.com applications (Call/Return)	& OPC C\&C View and Deployment View \\ \midrule
		Adventure Order Processing DB (executes on srv-dbopc) & The Order Processing Component uses this for consumer account data, consumer purchases and external invoices (R/W) & Consumer Website (mostly read for authentication, write only at account creation and update)
\newline Social features (read---need to characterize workload)
\newline Cross-sell (read---once per purchase) & OPC C\&C View and Deployment View
\\ \midrule
		Consumer UI (executes on srv-web1 and srv-web2) & Primary shop and purchase workflows. & Social Features---insert new content for tagging, sharing, etc.
\newline Cross-sell--- insert cross-sell features.	& Top Level Uses View and Install View and Deployment View
\\ \bottomrule
	\end{tabular} 
\end{center}

\clearpage

Now you are going to use the models provided in the view of the AB system to answer some questions. We have a few questions related to each of the new capabilities. Obviously, in a real integration project, there would be a multitude of questions: Here, we attempt to cover a small sample of these questions to assess the utility of the models.

In each answer, please note which models you consulted to make your decision. If you cannot find sufficient information in the models to make your decision, please state this (these questions were formulated without consulting the models).

First, let's consider our approach to achieving the first new capability: adding social features to the AB system. These social features will create runtime dependencies between the AB system and the Social Travel system.

\begin{center}
	\small
	\begin{tabular}{L{6cm} L{6cm}} \toprule
		\textbf{Question}&\textbf{Your Answer} \\ \midrule
		1.	The AB system has 7x24 availability. The Social Travel systems have had recent outages, and there is ongoing work to improve availability. Which AB stakeholders do you need to engage with to understand the AB availability requirements?&  \\ \midrule
		2.	You need to expose the social features in the AB web user interface, which uses gwt (Google Web Toolkit). The Social Travel system uses V2.7.0 of the gwt. Which version does AB use? &  \\ \midrule
		3.	Is there an external programmatic interface to access the User Account and Purchase Order data elements? &  \\ \bottomrule
	\end{tabular} 
\end{center}

Next, let's consider the second new capability, which requires us to replace the payment processing interface on the AB system. AB calls this interface the "Bank" interface.

\begin{center}
	\small
	\begin{tabular}{L{6cm} L{6cm}} \toprule
		\textbf{Question}&\textbf{Your Answer} \\ \midrule
		4.	The inputs to the new payment processing interface are: Card Type, Card Number, Card Expiration, and Card Security Code/CCV. Can the current AB system provide all of these elements?&  \\ \midrule
		5.	Using the new payment processing interface may impact the purchase completion latency of the AB system. Which stakeholders do you need to consult with about the performance requirements? &  \\ \bottomrule
	\end{tabular} 
\end{center}

\clearpage

Now, let's consider the third new capability, which adds cross-selling to the AB purchase user interface. Our approach is to have the AB system make a request to a Social Travel system when the purchase order has been placed, the necessary details about the travel order (e.g., traveler's gender, destination, date, activity type, etc.) and the Social Travel cross-sell business logic will return a list of five items to offer to the traveler.

\begin{center}
	\small
	\begin{tabular}{L{6cm} L{6cm}} \toprule
		\textbf{Question}&\textbf{Your Answer} \\ \midrule
		6.	Does the AB system have existing request-response interfaces with external systems? If so, what protocols/technologies are used for these interfaces?&  \\ \midrule
		7.	Is the traveler's gender available in the AB system?  &  \\ \midrule
		8.	Does an AB purchase order contain the activities that are included in the adventure? &  \\ \bottomrule
	\end{tabular} 
\end{center}

Finally, let's go back to the three questions that you listed at the very start of the exercise. Try to use the models to answer these questions---if one of your questions duplicates one of the previous questions, you can just note that here.
In each answer, please note which models you consulted to make your decision. If you cannot find sufficient information in the models to make your decision, please state this.

\begin{center}
	\small
	\begin{tabular}{L{6cm} L{6cm}} \toprule
		\textbf{Question}&\textbf{Your Answer} \\ \midrule
		Your first question (There is no need to copy the question again here, unless you find that helpful)&  \\ \midrule
		Your second question  &  \\ \midrule
		Your third question &  \\ \bottomrule
	\end{tabular} 
\end{center}

Two wrap-up questions:

\begin{center}
	\small
	\begin{tabular}{L{6cm} L{6cm}} \toprule
		\textbf{Question}&\textbf{Your Answer} \\ \midrule
		How much time did you spend on this exercise?&  \\ \bottomrule
	\end{tabular} 
\end{center}

Do you have any comments on the exercise? Did the role-playing questions represent the types of questions that an architect responsible for this integration might ask? Did you notice any gaps in the coverage of the models that was not exposed by the questions? Do you have any comments about the utility of the models in this viewpoint?

\textbf{Your Answer:}

\begin{center}
	\textbf{This is the end of the exercise. Thank you for your contribution. We will share the results with you as soon as we complete the analysis.}
\end{center}

\clearpage
\renewcommand\bibname{References}
\bibliographystyle{elsarticle-num}
\bibliography{sos-viewpoint}

\end{document}